\documentclass[runningheads]{llncs}

\usepackage[T1]{fontenc}
\usepackage{graphicx}
\usepackage{diagbox}
\usepackage{amsfonts}
\usepackage{hyperref}
\usepackage{amsmath}
\usepackage[capitalise]{cleveref}
\usepackage{tabularray}
\usepackage{multirow}
\usepackage{enumitem}

\usepackage{xcolor}
\hypersetup{
    colorlinks,
    linkcolor={blue!50!black},
    citecolor={blue!50!black},
    urlcolor={blue!50!black}
}
\usepackage{color}

\usepackage{bbm}

\begin{document}

\title{Unsupervised Latent Stain Adaptation\\ for Computational Pathology}

%
\author{Daniel Reisenb\"uchler\inst{1,}\textsuperscript{$\star$,}\textsuperscript{@}
\and 
Lucas Luttner\inst{1,}\textsuperscript{$\star$}\and 
Nadine S. Schaadt\inst{2} \and \\
Friedrich Feuerhake\inst{2} \and 
Dorit Merhof\inst{1,3} 
}
\authorrunning{D. Reisenb\"uchler et al.}
%
\institute{Institute of Image Analysis and Computer Vision, University of Regensburg, Regensburg, Germany \and
Institute of Pathology, Hannover Medical School, Hannover, Germany \and
Fraunhofer Institute for Digital Medicine MEVIS, Bremen, Germany\\
\textsuperscript{$\star$}These authors contributed equally to this work.\\
\textsuperscript{@}Correspondence: \href{mailto:francescopaolo.casale@helmholtz-munich.de}{daniel.reisenbuechler@informatik.uni-regensburg.de}\\
} 
\maketitle
\begin{abstract}
In computational pathology, deep learning (DL) models for tasks such as segmentation or tissue classification are known to suffer from domain shifts due to different staining techniques. Stain adaptation aims to reduce the generalization error between different stains by training a model on source stains that generalizes to target stains. Despite the abundance of target stain data, a key challenge is the lack of annotations. To address this, we propose a joint training between artificially labeled and unlabeled data including all available stained images called Unsupervised Latent Stain Adaptation (ULSA). Our method uses stain translation to enrich labeled source images with synthetic target images in order to increase the supervised signals. Moreover, we leverage unlabeled target stain images using stain-invariant feature consistency learning. With ULSA we present a semi-supervised strategy for efficient stain adaptation without access to annotated target stain data. Remarkably, ULSA is task agnostic in patch-level analysis for whole slide images (WSIs). Through extensive evaluation on external datasets, we demonstrate that ULSA achieves state-of-the-art (SOTA) performance in kidney tissue segmentation and breast cancer classification across a spectrum of staining variations. Our findings suggest that ULSA is an important framework for stain adaptation in computational pathology.
\end{abstract}
\section{Introduction}
Recent advances in DL for computational pathology have shown promising results for a wide range of applications, from cancer and biomarker detection~\cite{LAMIL} to tissue structure segmentation~\cite{nassim_dataset}. However, large-scale studies have shown that the effectiveness of DL techniques in histology is highly dependent on the availability of labeled data~\cite{cancer_cell_msi}. Despite its theoretical promise, acquiring a sufficient number of expert annotations remains challenging. In the realm of digital pathology, image datasets often consist of sequential slides stained with various techniques, each providing different insights into the same region of interest. Despite variations in staining protocols, these slides often share a significant amount of consistent information. However, expert annotations may be available for one type of staining but may be lacking for others, which are often accessible in large quantities without labels. Generating expert annotations for multiple staining techniques for the same analysis tasks would be extremely time-consuming. In the era of foundation models~\cite{found_models}, we also prefer generalized DL models that are robust to data shifts instead of domain experts.
In this paper, we question how to tailor a DL model trained for a specific task to handle variations in staining within the distribution of target stains, for which no annotations are available. This can be accomplished by incorporating unlabeled data during the training phase. The aspect of stain adaptation across different inter-staining techniques has not been sufficiently explored so far. Despite efforts to develop stain-to-stain translation techniques, their effectiveness is typically evaluated either visually by experts or through translation metrics~\cite{stain_bench}. 
Previous research has not focused on directly incorporating unlabeled target stain images into the training process, but only used them for translation~\cite{nassim,friedrich_feuerhake}.
\noindent Here, we present ULSA, a semi-supervised strategy designed for joint training of all staining data for the first time. We introduce a framework that integrates unlabeled target stain images into supervised training by maintaining the supervised learning signal for synthetic target stainings generated through Cycle GAN (cGAN) inference~\cite{cGAN}. Feature-wise stain-adaptation enables using unlabeled target data and enforces feature consistency across stains. By combining these key ingredients, we propose a new method for efficient stain adaptation that outperforms current SOTA approaches.
Our novelties can be summarized as 
\begin{itemize}[leftmargin=*,label={}]
  \item \textbf{(1) Unsupervised Stain Adaptation.} ULSA leverages target stain data in a supervised and unsupervised manner using only annotated source stains. We propose a framework for training stain-invariant models for digital pathology.
  \item \textbf{(2) Feature Consistency Learning.} We maximize cosine-similarity between hierarchical features across stains to achieve stain-invariance on feature level.
  \item \textbf{(3) Task Agnostic Framework.} ULSA is applicable for classification and segmentation training of stain-invariant models.
  \item \textbf{(4) Outperforming SOTA.} Our approach outperforms methods from stain-translation, DL based augmentation and semi-supervised learning slightly for source stains and by a large margin for target stains. Our approach needs only 10\% of labels to achieve the same performance as SOTA trained on all data.
\end{itemize}
\section{Method}
\begin{figure} [t!]
    \centering
    \includegraphics[width=\textwidth]{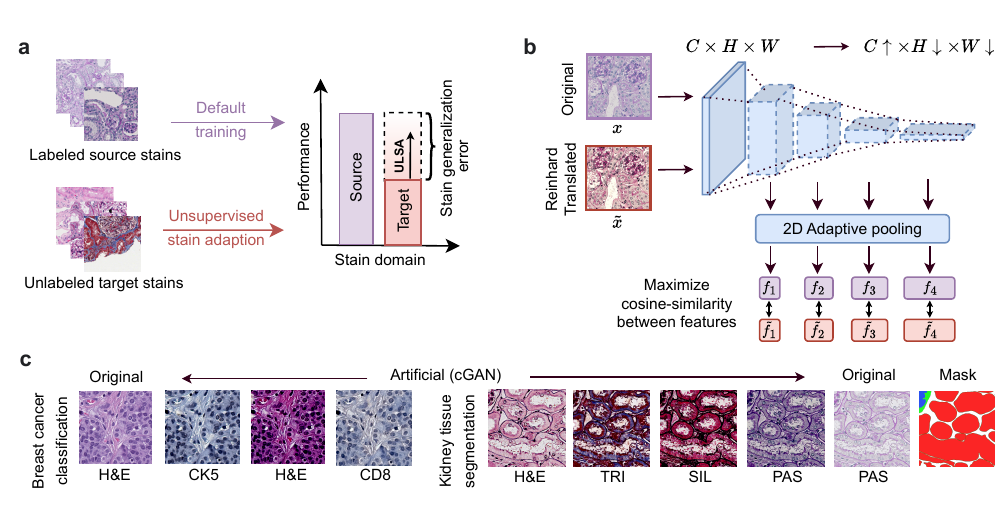}
    \caption{(\textbf{a}). Problem statement of unsupervised stain adaptation. (\textbf{b}). Stain-invariant feature consistency learning. (\textbf{c}). Artificial images generated by cGAN.} 
    \label{fig:fig_2}
\end{figure}

\begin{figure}[]
    \centering
    \includegraphics[width=\textwidth]{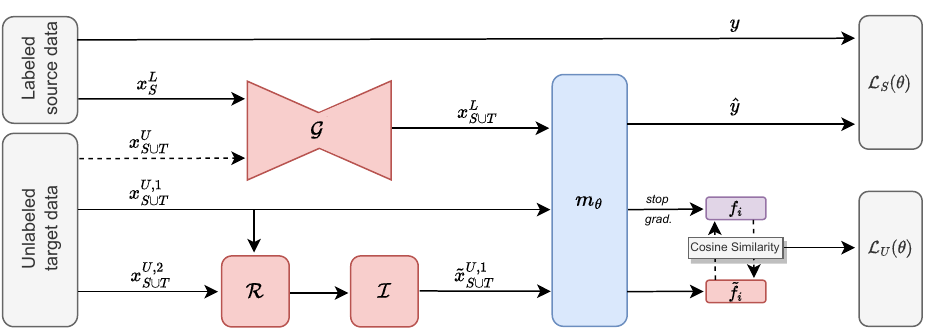}
    \caption{ULSA model. Labeled source stains are translated into synthetic target stain data to obtain supervision for image-wise stain-invariance. We extract features for a real and stain translated noised image of the target stain data, where we maximize cosine similarity to achieve unsupervised feature-wise stain-invariance in latent space.} 
    \label{fig:fig_framework}
\end{figure}

Stain adaptation aims to minimize the generalization error in task performance between a source stain $s\in S$ and a target stain $t\in T$ (Fig.~\ref{fig:fig_2}a). In particular, a parameterized model $m_{\theta}$ trained on labeled source staining data $x_{S}^{L}$ should ideally maintain task performance on other unlabeled target stains $x_{T}^{U}$ where no labels are available. We propose to address this challenge by incorporating unlabeled target stainings through (\textbf{\textsc{i}}) a cGAN model to augment labeled images into target stains that inherit the same annotation, (\textbf{\textsc{ii}}) unsupervised stain adaptation to jointly train on all stains with supervised and unsupervised objectives including all stains, followed by (\textbf{\textsc{iii}}) stain-invariant feature consistency learning (FCL) by unsupervised matching of latent representations between stains. The overall method is outlined in Fig.~\ref{fig:fig_framework}.
\subsubsection{(\textsc{i}) cGAN stain augmentation.} We pretrained and used cGANs, which we define as stain translation function $\mathcal{G}(x_{S}^{L})=x_{S \cup T}^L$ to synthetically augment source training data into target stainings. This process is structure preservering~\cite{cGAN}, thus each target stain image inherits the label corresponding to the associated source image $x_{S}^{L}$ used for translation. 
Fig.~\ref{fig:fig_2}c shows exemplary translation results. This strategy aims to increase the labeled training dataset and thus the supervising signal to achieve stain-invariance on prediction level. 
\subsubsection{(\textsc{ii}) Unsupervised stain adaptation.}
Since labeled data are given only in source staining $x^{L}\sim p_{S}(x^{L})$, we would ideally desire a mapping from source to target samples. We approximate this mapping by using $|S| \cdot |T|$ distinct cGAN augmenter to obtain additional labeled target samples $x^{L} \sim \hat{p}_{t\in T}(x^{L}\mid x^{U})$. With $|\cdot|$ we denote the cardinality of a set of stains. Note that all inferred samples $x^{L}$ inherit the label $y$ associated with $x^{L}$. Thus, we approximate $p_{S\cup T}$ by a mixture
\[
\hat{p}_{S \cup T}\left(x^{L}\right) = \frac{1}{|S|+|T|} \left( \sum_{s\in S}p_{s}\left(x^{L}\right) + \sum_{t\in T} \hat{p}_{t}\left(x^{L}\right)\right).
\]
In addition, we leverage unlabeled data by unsupervised learning. Given an unlabeled image $x_{t\in T}^{U, 1}$, we translate it to $\tilde{x}_{t\in T}^{U, 1}=\mathcal{I}\left(\mathcal{R}\left(x_{t\in T}^{U, 1}\mid x_{t\in T}^{U,2}\right)\right)$ by Reinhard translation $\mathcal{R}$ and noise injection $\mathcal{I}$. Note that $\mathcal{R}$ receives another unlabeled random sampled image $x_{t\in T}^{U,2}$ from target stains $t\in T$ as reference for subsequent translations. Fig.~\ref{fig:fig_2}b include example inputs. We used Reinhard normalization as Macenkos method has much higher runtime leading to computational overhead, see Supplementary Material (SM). Finally, we enforce the model to embed images invariant to stain translations by maximizing cosine similarity in unsupervised loss $\mathcal{L}_{U}$, see (\textsc{iii}). For this reason, we used light Gaussian blurring for noise injection, see SM for other choices. In summary, we define our objective as
\[
\min_{\theta}\mathcal{L}\left( \theta \mid x^{L}, y, x^{U}\right) = 
\underset{x^{L}\sim \hat{p}_{S \cup T}\left(x^{L}\right)}{\mathbb{E}}
\left[ \mathcal{L}_{S}\left(\theta \mid x^{L}, y\right)\right] + \lambda \underset{x^{U}\sim p_{S \cup T}\left(x^{U}\right)}{\mathbb{E}}
\left[ \mathcal{L}_{U}\left(\theta \mid x^{U}\right)\right]
\]
where $\mathcal{L}_{S}$ is a supervised loss. With minimizing $\mathcal{L}_{S}$ using labeled images of synthetic target stains, we aim for stain-adaptation on prediction level. We use multi-class and binary cross-entropy loss for segmentation and classification, respectively. We set equal weight $\lambda=1$, obtained by performing a grid search as described in SM. In each iteration we compute $\mathcal{L}_S$ on a batch of labeled data and compute feature consistency with $\mathcal{L}_{U}$ on a batch of unlabeled data. Next, we detail the unsupervised loss term associated with feature consistency learning.
\subsubsection{(\textsc{iii}) Stain-invariant feature consistency learning.} For unsupervised learning, we aim to calculate similar feature representations for an input $x=x_{t\in T}^{U, 1}$ and a stain translated noised version $\tilde{x}=\tilde{x}_{t\in T}^{U, 1}$. To enforce this for every downsampling block in a model, we apply non-parametric 2D adaptive average pooling. In Einstein notation, this tensor operation leads to $B,C_{i},H_{i},W_{i}\rightarrow B,C_{i}$ where $B$, $C$, $H$, $W$ and $i$ refer to batch-size, channels, height, width and the model block, respectively (Fig.~\ref{fig:fig_2}b). We aim to maximize the cosine similarity between all hierarchical features of a model $m_{\theta}$. In this way, we enforce feature similarity by updating the model such that extracted image features become similar in latent space. We define features as model outputs up to block $i$ after 2D adaptive pooling, $f_{\bar{\theta}, i}=m_{\bar{\theta}, i}(x)$, $\tilde{f}_{\theta, i}=m_{\theta, i}(\tilde{x})$. For brevity, index $\theta,i$ also refers to parameter optimization from input layer until block $i$. Following the literature \cite{miyato2018virtual,uda}, we do not backpropagate gradients for non-augmented input, denoted with $\bar{\theta}$. Hence, we define our unsupervised objective $\mathcal{L}_{u}$ as 
\[
\mathcal{L}_{U}(\theta) = - \frac{1}{b} \sum_{i\in b} \frac{ \mathbf{f}_{\bar{\theta},i}\cdot \tilde{\mathbf{f}}_{\theta, i}} {\lVert \mathbf{f}_{\bar{\theta},i} \rVert_{2} \cdot \lVert \tilde{\mathbf{f}}_{\theta, i}\rVert_{2}},
\]
where $\lVert \cdot \rVert_{2}$ denotes the Euclidean norm and $b$ the number of downsampling blocks. With minimizing $\mathcal{L}_{U}$, we enforce stain-invariance at the feature level. 
\section{Experiments}
We applied our proposed method to kidney tissue segmentation and cancer classification. We measured the performance using the dice score and the area under the receiver operating characteristic curve (AUROC), respectively. Next, we describe our datasets and more information about statistics can be found in SM.  
\subsection{Datasets and Comparable Methods}
For slide tiling, we used a modified version of the CLAM preprocessing pipeline~\cite{clam_prepro} and manually selected color thresholds as different stains require adjustments in tissue detection. All images were processed at a size of $512\times512$ px.
\subsubsection{Kidney segmentation datasets.} Our internal annotated train $n_{train}^{PAS}=2{,}100$ and validation $n_{val}^{PAS}=160$ datasets consist of annotated patches extracted from periodic acid-Schiff (PAS) stained WSIs with $20\times$ magnification. The annotation masks contain the classes tubule, glomerulus, glomerular tuft, artery, arterial lumen, and vein. 
For external testing we included annotated glomerulus images from PAS stained WSIs provided by the HuBMAP consortium~\cite{hubmap}. After processing the raw data we obtained $n_{ext, hubmap}^{PAS}=2{,}670$ samples. Moreover, we included test data from the NEPTUNE~\cite{neptune} study, containing $n_{ext, neptune}^{HE}=402$, $n_{ext, neptune}^{PAS}=1{,}176$, $n_{ext, neptune}^{SIL}=688$ Silver (SIL) and $n_{ext, neptune}^{TRI}=817$ Trichome (TRI) stain samples. The classes glomerulus, glomerular tuft, artery and tubule were annotated and images were taken from WSIs with $40\times$ magnification and rescaled to $20\times$ magnification. We further processed unlabeled WSIs at $20\times$ magnification from the KPMP database~\cite{kpmp}. We obtained $n_{unlabeled}^{HE}=385{,}670$, $n_{unlabeled}^{PAS}=409{,}554$, $n_{unlabeled}^{SIL}=538{,}412$, $n_{unlabeled}^{TRI}=415{,}822$ tiles.  
\subsubsection{Breast cancer classification datasets.} Our cancer classification datasets contain $n_{train}^{HE}=1{,}950$, $n_{val}^{HE}=287$ images obtained from HE stained WSIs and tiled at $40\times$ magnification level. Our test set contains Cytokeratin (CK5) and Cluster of differentiation (CD8) stains with $n_{test}^{CK5}=1{,}000$ and $n_{test}^{CD8}=1{,}000$ samples. Sample sizes and binary label ratios were equalized for test sets after preprocessing and excess data were held out of experiments to avoid data leakage at patient level. Additionally we have unlabeled datasets containing $n_{unlabeled}^{HE}=117{,}369$, $n_{unlabeled}^{CK5}=41{,}735$, $n_{unlabeled}^{CD8}=52{,}492$ samples. We split the data at slide-level so that each patient appears exclusively in one dataset.
\subsubsection{Comparable methods.} We selected comparable methods from the domains of stain translation, unsupervised augmentation, and semi-supervised consistency training. \textbf{(1) Baseline}. For comparison, we include a naive approach where we train a model on source stains without access to target stains, thus obtaining a lower bound for stain adaptation.
\textbf{(2) Reinhard}. Reinhard's method is a stain normalization technique that adjusts the color appearance of histopathology images by aligning them with a reference color space~\cite{reinhard}. \textbf{(3) Macenko}. Macenko's method~\cite{macenko} is a stain normalization technique that standardizes the color appearance of images by mapping them to a reference color space. It has been shown that this method provides the best performance across various colorization techniques for downstream tasks~\cite{friedrich_feuerhake}. \textbf{(4) cGAN Augmentation.} This method generates synthetic images with diverse staining variations in order to translate between stainings and train stain-invariant models~\cite{nassim}. \textbf{(5) FixMatch}. A semi-supervised learning approach that combines labeled and unlabeled data by enforcing consistency between predictions made on unlabeled samples based on confidence scores~\cite{fixmatch}. \textbf{(6) Unsupervised Data Augmentation (UDA)}. UDA is a semi-supervised learning technique designed for classification tasks that leverages augmentations on unlabeled data to improve model performance by minimizing probability distributions between two versions of an image. Originally proposed for natural images~\cite{uda} it has been adapted to histology images~\cite{uda_pathology}. 
\subsection{Implementation}\label{implementation}
\subsubsection{cGAN augmentation.} We initially started with hyperparameters following the literature~\cite{nassim} and performed a grid search with details noted in SM. By visually evaluating the stain-translation results, we selected a model trained with $300$ epochs and a learning rate of $1.5e-4$. For each stain translation we used $10,000$ unlabeled images from KPMP. For source stains, we used unlabeled HE data for cancer classification, and our labeled PAS stained training set for segmentation. Each stain translation training and inference task took about 8 hours. 
\subsubsection{Segmentation and classification.} For all experiments we used a ResNet-50 as classification model and encoder model for U-Net in segmentation. We tuned hyperparameters on validation sets and set a learning rate decay of $1e-10$ and early stopping for 5 and 10 consecutive epochs with no decrease in validation loss. We employed AdamW as optimizer and set the initial learning rate to $1e-04$ and the weight decay to $1e-05$.  Images were resized to a scale of $224\times224$ px. We used a total batch-size of 128 for all experiments. For semi-supervised learning we batched data to 32 and 96 for labeled and unlabeled data. All models were initialized with ImageNet weights and experiments were run on a single Nvidia A100 GPU. We measured a maximum training time of 10 and 7 hours across methods (except Macenko) for segmentation and classification, respectively.
\section{Results and Discussion}
In this section, we report the results and analyze the number of labels required for stain adaptation in kidney tissue segmentation and breast cancer classification. 
\begin{table}[ht]
\caption{Dice and AUROC scores for segmentation and classification tasks, respectively. We report the mean and standard deviation across three consecutive runs. All methods (except Baseline) access unlabeled target stains in the trainings phase.}\label{tab:results}
\centering
\renewcommand{\arraystretch}{1.1}
\begin{tabular}{c|ccccc|ccc}
\hline
\multirow{3}{*}{Method}                                   & \multicolumn{5}{c|}{Segmentation}                                                                                                                                                                                                                                                     & \multicolumn{3}{c}{Classification}                                                                                                                                    \\ \cline{2-9} 
                                                          & \multicolumn{1}{c|}{Intra-stain}                      & \multicolumn{4}{c|}{Inter-stain}                                                                                                                                                                                              & \multicolumn{3}{c}{Inter-stain}                                                                                                                                       \\
                                                          & \multicolumn{1}{c|}{PAS}                              & TRI                                                   & HE                                                    & SIL                                                   & \textbf{Overall}                                                & CK5                                                   & CD8                                                   & \textbf{Overall}                                                \\ \hline
Baseline                                                  & \begin{tabular}[c]{@{}c@{}}87.4\\ (0.66)\end{tabular} & \begin{tabular}[c]{@{}c@{}}59.4\\ (4.19)\end{tabular} & \begin{tabular}[c]{@{}c@{}}43.2\\ (4.01)\end{tabular} & \begin{tabular}[c]{@{}c@{}}76.4\\ (1.67)\end{tabular} & \begin{tabular}[c]{@{}c@{}}62.1\\ (3.24)\end{tabular} & \begin{tabular}[c]{@{}c@{}}86.6\\ (9.61)\end{tabular} & \begin{tabular}[c]{@{}c@{}}90.5\\ (7.10)\end{tabular} & \begin{tabular}[c]{@{}c@{}}88.6\\ (8.35)\end{tabular} \\
Reinhard~\cite{reinhard}            & \begin{tabular}[c]{@{}c@{}}87.3\\ (0.50)\end{tabular} & \begin{tabular}[c]{@{}c@{}}64.8\\ (6.50)\end{tabular} & \begin{tabular}[c]{@{}c@{}}40.2\\ (3.23)\end{tabular} & \begin{tabular}[c]{@{}c@{}}77.9\\ (2.57)\end{tabular} & \begin{tabular}[c]{@{}c@{}}64.3\\ (4.39)\end{tabular} & \begin{tabular}[c]{@{}c@{}}89.8\\ (3.65)\end{tabular} & \begin{tabular}[c]{@{}c@{}}94.1\\ (1.48)\end{tabular} & \begin{tabular}[c]{@{}c@{}}91.9\\ (2.32)\end{tabular} \\
Macenko~\cite{friedrich_feuerhake} & \begin{tabular}[c]{@{}c@{}}85.0\\ (0.80)\end{tabular} & \begin{tabular}[c]{@{}c@{}}71.6\\ (2.60)\end{tabular} & \begin{tabular}[c]{@{}c@{}}48.3\\ (4.70)\end{tabular} & \begin{tabular}[c]{@{}c@{}}81.6\\ (0.50)\end{tabular} & \begin{tabular}[c]{@{}c@{}}70.3\\ (2.29)\end{tabular} & \begin{tabular}[c]{@{}c@{}}89.5\\ (2.90)\end{tabular} & \begin{tabular}[c]{@{}c@{}}93.4\\ (2.06)\end{tabular} & \begin{tabular}[c]{@{}c@{}}91.4\\ (2.48)\end{tabular} \\
cGAN~\cite{nassim}                  & \begin{tabular}[c]{@{}c@{}}84.2\\ (0.92)\end{tabular} & \begin{tabular}[c]{@{}c@{}}69.9\\ (1.44)\end{tabular} & \begin{tabular}[c]{@{}c@{}}46.0\\ (4.71)\end{tabular} & \begin{tabular}[c]{@{}c@{}}79.0\\ (1.63)\end{tabular} & \begin{tabular}[c]{@{}c@{}}68.1\\ (2.18)\end{tabular} & \begin{tabular}[c]{@{}c@{}}87.1\\ (6.24)\end{tabular} & \begin{tabular}[c]{@{}c@{}}88.0\\ (3.32)\end{tabular} & \begin{tabular}[c]{@{}c@{}}87.5\\ (3.67)\end{tabular} \\
FixMatch~\cite{fixmatch}            & \begin{tabular}[c]{@{}c@{}}87.2\\ (0.51)\end{tabular} & \begin{tabular}[c]{@{}c@{}}64.8\\ (2.67)\end{tabular} & \begin{tabular}[c]{@{}c@{}}40.3\\ (7.28)\end{tabular} & \begin{tabular}[c]{@{}c@{}}78.2\\ (1.02)\end{tabular} & \begin{tabular}[c]{@{}c@{}}64.5\\ (3.05)\end{tabular} & \begin{tabular}[c]{@{}c@{}}88.3\\ (2.25)\end{tabular} & \begin{tabular}[c]{@{}c@{}}94.3\\ (1.37)\end{tabular} & \begin{tabular}[c]{@{}c@{}}91.3\\ (1.78)\end{tabular} \\
UDA~\cite{uda_pathology}           & \begin{tabular}[c]{@{}c@{}}87.2\\ (0.54)\end{tabular} & \begin{tabular}[c]{@{}c@{}}64.9\\ (1.84)\end{tabular} & \begin{tabular}[c]{@{}c@{}}45.8\\ (2.49)\end{tabular} & \begin{tabular}[c]{@{}c@{}}77.5\\ (0.60)\end{tabular} & \begin{tabular}[c]{@{}c@{}}65.4\\ (1.53)\end{tabular} & \begin{tabular}[c]{@{}c@{}}89.7\\ (2.19)\end{tabular} & \begin{tabular}[c]{@{}c@{}}92.8\\ (1.97)\end{tabular} & \begin{tabular}[c]{@{}c@{}}91.3\\ (1.47)\end{tabular} \\ \hline 
Ours                                                      & \begin{tabular}[c]{@{}c@{}}\textbf{87.9}\\ (0.33)\end{tabular} & \begin{tabular}[c]{@{}c@{}}\textbf{74.1}\\ (0.94)\end{tabular} & \begin{tabular}[c]{@{}c@{}}\textbf{53.6}\\ (1.46)\end{tabular} & \begin{tabular}[c]{@{}c@{}}\textbf{81.8}\\ (0.61)\end{tabular} & \begin{tabular}[c]{@{}c@{}}\textbf{72.6}\\ (0.93)\end{tabular} & \begin{tabular}[c]{@{}c@{}}\textbf{91.6}\\ (4.63)\end{tabular} & \begin{tabular}[c]{@{}c@{}}\textbf{94.6}\\ (1.40)\end{tabular} & \begin{tabular}[c]{@{}c@{}}\textbf{93.1}\\ (3.01)\end{tabular}
\end{tabular}
\end{table}
\subsubsection{Measuring stain adaptation.} In the area of stain adaptation, we aim to at least maintain performance on source stains (Intra-stain, Tab.~\ref{tab:results}) and maximize performance on target stains (Inter-stain, Tab.~\ref{tab:results}). For segmentation on source data our method is on par with other methods, and also shows slightly improved performance. Note that all other methods decrease their performance on source stains compared to the baseline. More importantly, we increase the Dice score for target stains (Inter-stain, Overall) by more than $10$ and $2.3$ compared to naive training and Macenko colorization, the best comparable method. Note, however, that although Macenko provides the second best stain adaptation result, source performance is reduced by $2.4$ and $2.9$ compared to baseline and ULSA. In the case of classification, we did not have further annotated intra-stain data for testing, but report results for target data. Our method increases the AUROC by 4.5 and 1.2 compared to the naive and best comparable methods. Overall, these results demonstrate efficient stain adaptation by increasing the target while not only maintaining but increasing source performance (Fig.~\ref{fig:fig_2}a). Additionally, we trained all methods on different fractions of labeled data and tested their performance on targets in segmentation (Fig.~\ref{fig:train_fac}a). Interestingly, ULSA shows great stain adaptation even on very little annotated data and is on par with the best comparable method (on full data) using only 10\% of labeled data.
\subsubsection{Ablation study.} 
\begin{figure}[]
\centering
\includegraphics[width=\textwidth, keepaspectratio]{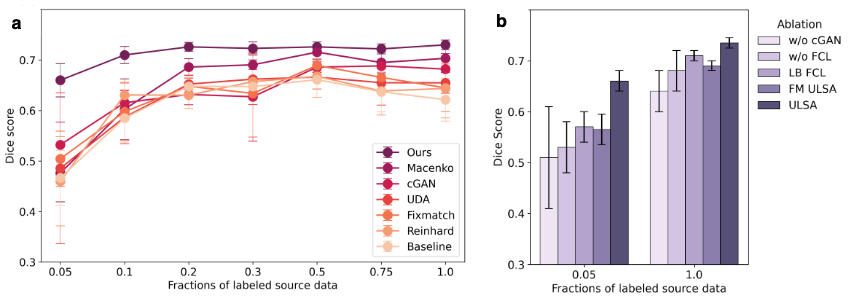}
\caption{(\textbf{a}). Performance for different fractions of labeled data. (\textbf{b}). Ablation study.}\label{fig:train_fac}
\end{figure}

We measured the influence of different components of our method in segmentation (Fig.~\ref{fig:train_fac}b). By dropping either cGAN or FCL components, overall target performance decreases. Using hierarchical features (ULSA) instead of features from the last block (LB FCL) increases the scores. We also initialized our ULSA method with pretrained weights from a foundation model for histology (FM ULSA), obtained by large-scale learning of HE images~\cite{retccl}. With this setup, we demonstrate that ImageNet pretrained weights yield better performance. This suggests that when building foundation models, stain variations should be considered to avoid catastrophic forgetting of learned morphologies.
\section{Conclusion}
We proposed ULSA, a novel SOTA strategy to reduce stain generalization errors in computational pathology tasks. Our semi-supervised learning strategy uses annotated data for both source and artificial target stains. In addition, we incorporate unlabeled data for stain-invariant feature consistency learning. Finally, joint optimization of supervised and unsupervised objectives enables efficient stain adaptation. We empirically demonstrated that ULSA training increases performance on unlabeled target stains in patch level segmentation and classification. This suggests that ULSA is a task agnostic framework. We further showed ULSA achieves efficient stain adaptation even in settings with scarce labels. A potential limitation of our approach is that even if the performance for unseen inter-stains is improved, augmentation may not translate the correct marker information from other stains. A possible example is immunohistochemical (IHC) staining, where specific immune cells are highlighted. This could potentially affect downstream applications in certain scenarios not seen in this study. Future work could compare other augmentation strategies such as HistAuGAN~\cite{HistAuGAN}.
\subsubsection{Acknowledgements.}
This work was supported by the German Research Foundation (Deutsche Forschungsgemeinschaf, DFG) under project number 445703531 and the German Federal Ministry of Education and Research (Bundesministerium für Bildung und Forschung, BMBF) under project numbers 01IS21067A and 01IS21067B. The authors gratefully acknowledge the computational and data resources provided by the Leibniz Supercomputing Centre (www.lrz.de).

\subsubsection{Disclosure of Interests.}
The authors declare that they have no conflicts of interest related to this work.
\clearpage
\bibliographystyle{splncs04}
\bibliography{article}
\end{document}


\title{Supplementary Material\\Unsupervised Latent Stain Adaptation\\ for Computational Pathology}

%
\author{Daniel Reisenb\"uchler\inst{1,}\textsuperscript{$\star$,}\textsuperscript{@}
\and 
Lucas Luttner\inst{1,}\textsuperscript{$\star$}\and 
Nadine S. Schaadt\inst{2} \and \\
Friedrich Feuerhake\inst{2} \and 
Dorit Merhof\inst{1,3} 
}
\authorrunning{D. Reisenb\"uchler et al.}
%
\institute{Institute of Image Analysis and Computer Vision, University of Regensburg, Regensburg, Germany \and
Institute of Pathology, Hannover Medical School, Hannover, Germany \and
Fraunhofer Institute for Digital Medicine MEVIS, Bremen, Germany\\
\textsuperscript{$\star$}These authors contributed equally to this work.\\
\textsuperscript{@}Correspondence: \href{mailto:francescopaolo.casale@helmholtz-munich.de}{daniel.reisenbuechler@informatik.uni-regensburg.de}\\
} 
\maketitle              

\section{Test Datasets}
The following tables serve as overview of the number patches used for each staining and tissue combination in the segmentation task. 

\begin{table}[]
\caption{Overview of the NEPTUNE dataset used for the segmentation experiments}
\centering
\begin{tabular}{ccc}
\hline
Staining             & Tissue class    & \#Images            \\ \hline
\multirow{4}{*}{PAS} & Glomerulus      & 329 \\ \cline{2-3} 
                     & Glomerular Tuft & 352                  \\ \cline{2-3} 
                     & Tubule          & 231                  \\ \cline{2-3} 
                     & Artery          & 264                  \\ \hline
\multirow{3}{*}{SIL} & Glomerulus      & 248                  \\ \cline{2-3} 
                     & Glomerular Tuft & 217                  \\ \cline{2-3} 
                     & Artery          & 223                  \\ \hline
\multirow{3}{*}{TRI} & Glomerulus      & 260                  \\ \cline{2-3} 
                     & Glomerular Tuft & 233                  \\ \cline{2-3} 
                     & Artery          & 324                  \\ \hline
H\&E                 & Artery          & 402                  \\ \hline
\end{tabular}
\end{table}

\begin{table}[]
\caption{Overview of the HuBMAP dataset used for the segmentation experiments}
\centering
\begin{tabular}{ccc}
\hline
Staining & Tissue class& \#Images\\ \hline
PAS      & Glomerulus   & 2670      \\ \hline
\end{tabular}
\end{table}

\section{Hyperparameter search}
All hyperparameter searches were performed via grid search on validation sets only. In the following we detail hyperparameter selections.

\subsubsection{cGAN.} We performed a careful selection of hyperparameters to ensure that the images were perfectly translated into the target stainings. The number of epochs for the adversarial model was searched within the interval $[200, 500]$ and set to $300$, while the learning rate was adjusted to $1.5e-4$, within the search range $[1e-3, 1e-5]$. The momentum term of Adam was set to $0.5$, within the interval $[0.01, 1]$. The buffer for storing artifical images was fixed at $50$ $[10, 200]$, and the batch size was fixed at $2$, which achieved the best results within the interval of $[1, 4]$. Finally, the number of unlabeled training data was set to $10 000$, within the interval $[1000, 50 000]$.

\subsubsection{ULSA.} We performed grid search for finding the weighting $\lambda$ between $\mathcal{L_S}$ and $\mathcal{L_U}$ in the range of $[0.3, 1.5]$ with step size $\Delta = 0.1$. We used the overall batch sizes between labeled $b_L$ and unlabeled samples $b_U$ with $b_{overall}=b_{L} + b_{U} = 128$ where we tried $b_{U} = \lambda b_{L}$ for different $\lambda = 1,2,3$. We tested several noise injection approaches including salt and pepper, gaussian blurring and gaussian noise. Best results were archived with gaussian blurring (kernel size: $(3,5)$, intensity: $(0.01, 0.4)$). Other augmentation methods like color jitter and random sharpness adjustments were tried, but did not show promising results. We further tried to replace Reinhard with Macenko, which was not possible, due to computational overload (ULSA with Reinhard:  7-10h, ULSA with Macenko: at least 48-60h). Also translating the stains offline and storing them locally would not be possible, because of the huge amount of images needed to store: $x^{U}= 1.749.458^{|T|}, |T|=[3,4]$. 

\subsubsection{Comparable methods.} \textit{Reinhard and Macenko.} For each image in the mini-batch we used a random target stained image as reference for transformation. Thus each image was translated multiple times into different target stains during training. \textit{UDA.} We used various combinations such as color jitter and gaussian blurring (also see augmentations for ULSA) for data augmentation in the semi-supervised part. Other augmentations lead to worst results. The batch size factor $\lambda$ for unlabeled data in the unsupervised data augmentation procedure was set to $3$ as proposed by the authors within the interval $[3,5]$. \textit{FixMatch.} We used a confidence threshold of $0.95$, which was the same the authors used for their implementations. All other parameters were obtained as in UDA.

\bibliographystyle{splncs04}